\documentstyle[prl,aps,epsf,multicol]{revtex}
\tighten
\begin{document}
\draft
\title{The classical counterpart of entanglement}
\author{Arul Lakshminarayan}
\address{Physical Research Laboratory,\\
Navrangpura, Ahmedabad,  380 009, India.}   
\maketitle
\newcommand{\newc}{\newcommand}
\newc{\beq}{\begin{equation}}
\newc{\eeq}{\end{equation}}
\newc{\kt}{\rangle}
\newc{\br}{\langle}
\newc{\beqa}{\begin{eqnarray}}
\newc{\eeqa}{\end{eqnarray}}
\newc{\pr}{\prime}
\newc{\longra}{\longrightarrow}

\begin{abstract}
We define and explore the classical counterpart of entanglement in
complete analogy with quantum mechanics. Using a basis independent
measure of entropy in the classical Hilbert space of densities that are
propagated by the Frobenius-Perron operator, we demonstrate that at short
times the quantum and classical entropies share identical power laws
and qualitative behaviors. 
\end{abstract}


\begin{multicols}{2}

Quantum entanglement has been studied with renewed vigor in the recent
past due to its relevance in quantum information theory
\cite{Peres}. It has been recognized as an unique quantum resource
that could be utilized for potentially powerful and qualitatively
different quantum computers \cite{Steane}.  It measures the degree of
non-separability between individual parts of a composite system. That
this non-separability may have apparently dramatic effects was
implicitly exploited in the Einstein, Podolsky, Rosen paradox. It
appears that much of the strangeness of quantum phenomena follows from
the existence of such entanglement.  A classical system, composed
perhaps of very many interacting parts, could have at all times a
unique state assigned to these parts, in terms of their phase-space
variables. However it is quite clear that a quantum state is not the
counterpart of a classical phase-space point; although both describe
the maximum possible information we can have of the system within the
framework of the respective theories. 

To arrive at the classical counterpart of entanglement we study a
classical measure of non-separability of individual subsystems.  In
the spirit of entanglement we do not allow for transformations that
may mix the states of the individual subsystems, i.e., the allowed
transformations must be local. A Hilbert space formulation of
classical mechanics is the natural starting point towards this
end. Recently, classical mechanics on the Hilbert space has been
studied to understand certain aspects of quantum-classical
correspondence, especially in the presence of classical chaos
\cite{Haake}. The principal object is the Frobenius-Perron (F-P) operator,
that is the exponentiation of the Liouville operator, which time
evolves phase-space densities and is the classical counterpart of the
quantum unitary propagator. States in the Hilbert space are allowed
phase-space densities, and the measure of separability is simply the
separability of evolving phase-space densities. Thus the machinery of
quantum entanglement, measuring entanglement in a complex Hilbert
space, may be applied directly within this space of classical
densities. Recent work, with the aim of bringing
into sharper focus the unique nature of quantum entanglement, has
explored the classical part of correlations from an information
theoretic point of view \cite{Class}.

Another related aspect is the issue of entropy production, and the
definition of entropy in irreversible processes
\cite{ChaosSPISSUE}. The usual definition of entropy in an ensemble
through the integral version of the Shannon entropy, runs into
conceptual difficulties. As is well known it is strictly a constant
and thus some sort of coarse-graining is needed to define a changing
entropy. Entropies are then defined through partially integrated out
densities, say through the Bogoliubov hierarchy\cite{Dorfman}. On the
other hand a von Neumann entropy of the classical densities, as we
define below, will be non-negative by construction, need not be a
constant of motion, and may be thought of as the most natural possible
coarse-graining. A related work in this context which explores
quantum-classical correspondence for open chaotic systems is in
\cite{Arjendu}.

Due to the difficulty in defining a measure of entanglement for
multipartite spaces, we consider a system having just two particles.
Quantum mechanically the system is described by states in a bipartite
Hilbert space and as is well known there exists a Schmidt decomposition
which if it has more than one non-zero term implies entanglement, and
the converse is also true \cite{Peres}. Now consider the classical
mechanics as described in the phase-space $(q_i,p_i)$, $i=1,2$. The
flow is generated by a Hamiltonian $H$ and the F-P operator ${\mathcal
P}$ is defined through:
\beq
{\mathcal P}(t) = \exp(-t \, \{\;\; ,H \});  \; \rho(q_i,p_i,t)=
{\mathcal P}(t)\, \rho(q_i,p_i,0).
\eeq 
The F-P operator is unitary. Rather than considering a Hilbert space
of $L^1$ functions, for simplicity we consider the space of $L^2$
functions. For instance we may define the evolution of the square-root
of a classical density. This also seems natural in view to our wanting
to do comparisons with quantum entanglement. Let $\{F_{m,n}(q,p)\}$ be
an orthonormal basis set in such a Hilbert space, which we will
write compactly as $\{ |m,n \kt \}$. We consider the case where there
is a denumerably infinite number of such basis functions labeled by
the pair $(m,n)$. A pair is used to facilitate defining a basis in
each Hilbert space corresponding to each phase-space variable. Thus we
can consider the classical Hilbert space as being an outer product of
four Hilbert spaces. However in the following we will consider it as a
product space of two Hilbert spaces, each corresponding to one degree
of freedom. 

Expanding any density $\rho$ in the four-dimensional phase space in
terms of these functions we may write:
\beq
|\rho \kt  = \sum a(m_1,n_1; m_2,n_2) |m_1,n_1 \kt \, |m_2,n_2 \kt,
\label{denvect}
\eeq
where the $a(;)$ are expansion coefficients. Now the ``reduced
density matrix'' is simply the partial trace:
\beq
\hat{\rho_1}=\mbox{Tr}_2(|\rho \kt \br \rho |),
\eeq
while the von Neumann entropy of this reduced density matrix is 
the entropy of classical entanglement and is given by:
\beq
\label{clent}
S_C=\mbox{Tr}_1(\hat{\rho_1} \log(\hat{\rho_1})).
\eeq
The analogy to the quantum definitions are evident and the 
existence of a Schmidt decomposition follows. The entropy $S_C$
can be zero or positive, if it is zero there is no entanglement, 
and the densities are separable into each degree of freedom, while if it 
is positive this is not possible. The basis independence of the
entropy $S_C$ follows from the two trace operations,  and 
thus it represents a physically meaningful quantity, which by its 
construction ought to qualify as the classical counterpart of
quantum entanglement.

We now provide evidence for this by means of an example. We choose two
coupled standard maps for the possibility of studying potentially
interesting connections between deterministic chaos and entanglement
\cite{Reichl}. In fact the motivation for the present work is derived
from a previous work \cite{ArulEnt} where we studied a similar system
which showed that classical chaos enhances quantum
entanglement. Furthermore we showed that some Schmidt vectors were
``scarred'' by projections of certain classical periodic orbits. A
study of coupled kicked tops has also shown that quantum entanglement
increases linearly in time at a rate which is the sum of the classical
positive Lyapunov exponents \cite{Sarben}. These are indications that quantum
entanglement is not entirely immune to the lure of the classical.

Consider the classical map defined on the four-torus $T^2 \times T^2$:
\beqa
q_i^ \prime &=&q_i+p_i \;\;\;(\mbox{mod} \;1) \nonumber \\
p_i^ \prime&=&p_i - \partial V /\partial q_i^{\pr} \;\;\;(\mbox{mod} \;1),
\label{clmap}
\eeqa
where $i=1,2$ and the potential $V$ is
\beqa
V= -\frac{K_1}{(2 \pi)^2} \cos(2 \pi q_1)
-\frac{K_2}{(2 \pi)^2} \cos(2 \pi q_2)+ \nonumber\\
\frac{b}{(2 \pi)^2} \cos(2 \pi q_1)
\cos(2 \pi q_2)
\eeqa
This is a symplectic transformation on the torus $T^4$ and may be
derived from a kicked Hamiltonian in the standard manner
\cite{Reichl}. This is somewhat different from the previously studied
four-dimensional generalizations of the standard map, and has been
chosen to maximize symmetries.  The parameter $b$ controls the
interaction between the two standard maps, while $K_i$ determine the
degree of chaos in the uncoupled limit.  The advantage in studying
such maps is our ability to write explicitly the relevant unitary
operators, both classical and quantal.

The classical unitary F-P operator over one iteration of the map is 
\beq
\label{fullFP}
{\mathcal P}= {\mathcal P}_1 \otimes {\mathcal P}_2 \; {\mathcal P}_b,
\eeq
where ${\mathcal P}_i$ is the F-P operator for the standard map on
$T^2$ and acts on the single-particle Hilbert spaces, while ${\mathcal
P}_b$ is the interaction operator on the entire space. We use as a
single-particle basis the Fourier decomposition:
\beq
F_{m,n}(q,p)=\exp\left(2 \pi i (m\, q \,+\, n \,p)\right).
\eeq
Then the matrix elements of the F-P operators are:
\beq
\br m_i^{\pr},n_i^{\pr}|{\mathcal P}_i | m_i,n_i \kt = 
\mbox{J}_{m_i^{\pr}-m_i}\left(K_i n_i^{\pr} \right) 
\delta_{n_i-m_i, \,n_i^{\pr}},
\eeq
and
\beqa
\br m_1^{\pr},n_1^{\pr};  m_2^{\pr},n_2^{\pr}|{\mathcal P}_b|
\br m_1,n_1;  m_2,n_2 \kt
&=& \nonumber \\
\mbox{J}_{l+}\left(\frac{b(n_1+n_2)}{2}\right)  
\mbox{J}_{l-}\left(\frac{b(n_1-n_2)}{2}\right)  
\delta_{n_1,\,n_1^{\pr}}\delta_{n_2,\,n_2^{\pr}},
\eeqa
where 
\[ l\pm\,=\, \frac{1}{2}\left( (m_1-m_1^{\pr}) \pm (m_2-m_2^{\pr}) \right),\]
and the delta functions are Kronecker deltas. The orders of the Bessel J
functions are restricted to the integers.

If the initial density is described by a vector (Eq. (\ref{denvect})) 
$|\rho(0) \kt$ then 
\beq
|\rho(T) \kt = {\mathcal P}^T \, |\rho(0)\kt
\eeq
is the density after a time $T$. The same may of course be
equivalently obtained by means of the Liouville equation.  The number
of classical components excited increases with time in general
including increasingly larger frequencies. This is indicative of the
fine structure that is being created in phase-space by the stretching
and folding mechanism. The frequencies involved appears to increase
exponentially in time for chaotic systems, while we may expect a
polynomial growth for integrable or near-integrable systems. The
Arnold cat map provides a completely chaotic exactly solvable model
where this exponential increase is readily seen \cite{Ford}, while the
free rotor, corresponding to the case $K_i=0$ for the standard map on
$T^2$ is an exactly solvable integrable model where a linear growth is
apparent. With the flow of probability moving inexorably outward, for
chaotic systems, the infinite dimensional Hilbert space of classical
mechanics provides a setting for an approach to equilibrium, with
every smooth density tending towards the invariant density in a weak
sense \cite{Fox}.

The quantization of the symplectic transformation in Eq.~(\ref{clmap})
is a finite unitary matrix on a product Hilbert space of
dimensionality $N^2$, and $N=1/h$, where $h$ is a scaled Planck
constant. The classical limit is the large $N$ limit.  The
quantization is straightforward as there exists a kicked Hamiltonian
generating the classical map \cite{Reichl}.  The quantum standard map on the
two-torus in the position representation is
\beqa
\label{2dqmap}
\br n_i^{\pr}|U_i|n_i \kt \,&=&\, \frac{\exp(-i \pi/4)}{\sqrt{N}}\,  \exp\left
( i N \frac{K_i}{2 \pi} 
\cos(\frac{2\pi}{N}(n_i+\frac{1}{2}))\right) \nonumber \\ &\times&
 \exp\left(\frac{\pi i }{N}  (n_i-n_i^{\prime})^2\right).
\eeqa
The position kets are labeled by $n\,=\, 0,N-1$ and the position
eigenvalues are $(n+1/2)/N$ while the momentum eigenvalues are
$m/N$, $m=0,\ldots,N-1$. The quantum phases have been chosen to maximize
symmetries, and $N$ is taken to be even. The
four-dimensional quantum map is but a simple extension:
\beq
\label{4dqmap}
U= U_1 \otimes U_2 \, U_b,
\eeq
which is the quantum equivalent of the classical F-P operator in
Eq.~(\ref{fullFP}) and the interaction operator is 
\beqa
&\br n_1^{\pr},n_2^{\pr}|U_b| n_1,n_2 \kt& = 
\exp \left[ -i N \frac{b}{2 \pi} \cos\left(\frac{2 \pi}{N}(n_1+\frac{1}{2})\right) \right.\nonumber \times \\ && \left.
\cos\left(\frac{2 \pi}{N}(n_2+\frac{1}{2})\right) \right] \delta_{n_1,n_1^{\pr}}
\delta_{n_2,n_2^{\pr}}.
\eeqa
$ U$ is a unitary matrix and will induce mixing between the two
subsystems. Some properties of such unitary matrices especially in 
connection with quantum entanglement has been studied 
previously \cite{ArulEnt}.

For the initial state a product state composed of coherent states
is a natural choice, as the corresponding initial  classical density 
becomes apparent. We choose 
\beq
|\psi(0)\kt = |00\kt \otimes |00 \kt,
\eeq
where $|00 \kt$ is a coherent state that is peaked at the classical
point $(q=0, p=0)$ \cite{Saraceno}, which also happens to be a
classical fixed point for the dynamics. The initial classical density
will be a circular (periodicised) gaussian of width $\sigma$ around
the origin.  We choose
\beq
a_0(m,n)= (8 \pi \sigma^2)^{1/2} \exp(-4 \pi^2 \sigma^2 (m^2+n^2)),
\eeq
to be  the Fourier components of such a  (approximate) periodicised gaussian.
The initial density's components are then specified by the outer product 
\beq
a_0(m_1,n_1;m_2,n_2)=a_0(m_1,n_1)\, a_0(m_2,n_2).
\eeq
For $\sigma \sim 1/\sqrt{N}$ the Husimi distribution of the quantum
state is similar to the (square of the) classical initial density. For
quantum-classical correspondence we require large $N$ or small
$\sigma$.

We time evolve the initial states, both quantum and classical, using
the respective propagators and calculate the entanglement in each.
Due to the infinite dimensionality of the classical Hilbert space and
the at least polynomially increasing frequency components, the
classical calculations in a truncated Hilbert space lead to rapid loss
of accuracy.  Thus this work, whose main intention is to compare the
classical entanglement as defined in Eq.~(\ref{clent}) with quantum
entanglement, merely touches the surface of interesting results. We
put $K_1=K_2=0$ and turn on the interaction parameter $b$ only
slightly.  Thus the uncoupled systems are free rotators, while the
coupled system is near-integrable. For sufficiently large values of the
interaction, full fledged chaos seems to develop, a case which we do not
discuss further here.

First we look at entanglement produced for very short times, as this
is numerically easily accessible. We study this entanglement as a
function of the interaction. In Fig.~\ref{entint} is shown
entanglement produced at times $1$ and $2$. 
{\narrowtext
\begin{figure}[h]
\epsfxsize=3.5in
\epsfysize=4in
\epsfbox{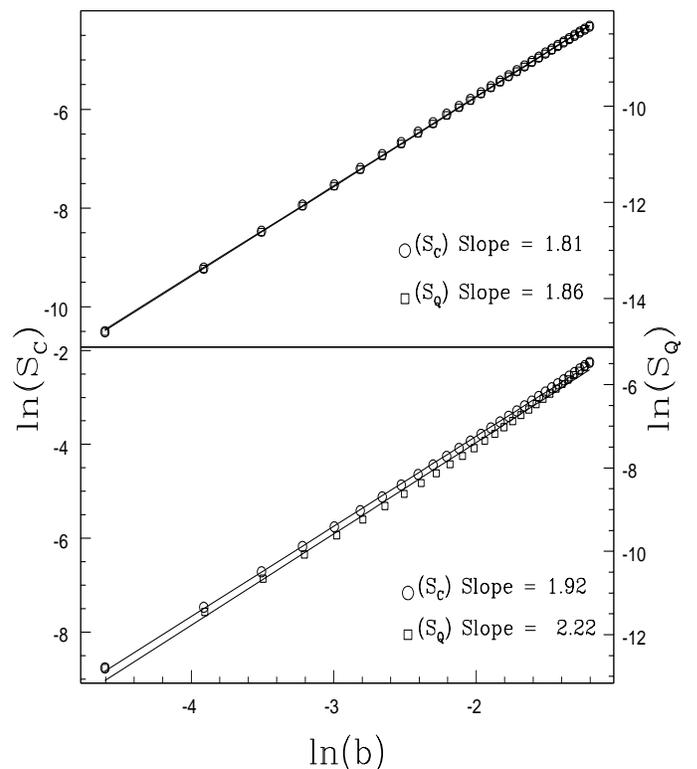}
\caption{Classical and quantum entanglement after time $T=1$ (top)
and $T=2$ (bottom) as a function of the interaction for the initial
state described in the text.  The quantum calculations correspond to
$N=50$, while the classical initial state has $\sigma=0.1$.}
\label{entint}
\end{figure}
}
\noindent We find
approximate power laws that are nearly the same for both classical and
quantum entanglement. The best correspondence is naturally at the
shortest time, $T=1$, when a power law of the form $S \sim b^{1.8}$
holds in both cases, while at time $2$ the exponents are not so
close. However at this time the power law seems to be approximate and
the deviations from the straight line is seen to have common features
in both the classical and quantum cases.  The $N$ or $\hbar$
dependence of the quantum entanglement does not seem to be a power law
at these short times and we do not pursue this here, thus we do not
compare the absolute values of the two entanglement.

In Fig.~\ref{enttime} we plot both the entanglements as a function of
time for a fixed value of the interaction. We see qualitatively
similar behaviour. The quantum entanglement tend to saturate at
higher values for larger $N$, while the classical entanglement seems
to be continuously increasing. This is however difficult to see
numerically due to increase of errors, for instance at the highest
time shown in the figure, $T=6$, the normalization deteriorated from
unity to about $.9948$. More elaborate work is needed to establish
the properties of classical entanglement.
{\narrowtext
\begin{figure}[h]
\epsfxsize=3.5in
\epsfysize=3in
\epsfbox{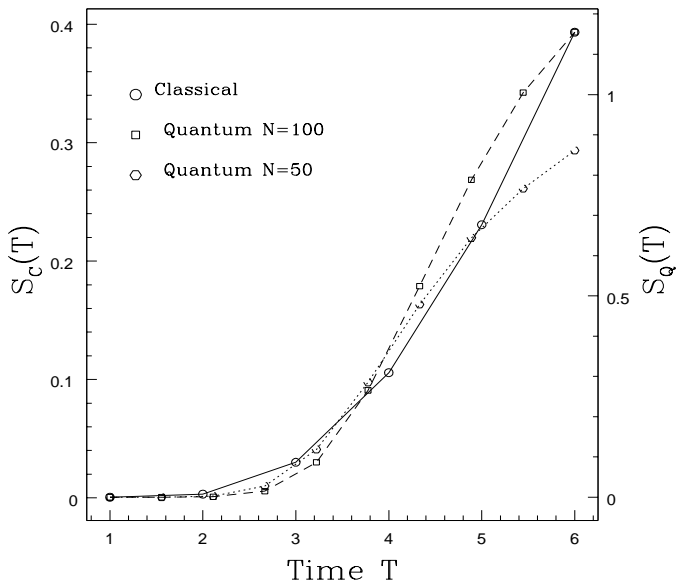}
\caption{Classical and quantum entanglement as a function of time for
the same initial state as in Fig.~\ref{entint}. The interaction $b=0.05$.}
\label{enttime}
\end{figure}
} We focus finally on an interesting regime where both the subsystems
are chaotic, while the interaction is weak. In this case a linear
quantum entropy increase was found, with the rate being proportional
to the sum of positive Lyapunov exponents \cite{Arjendu,Sarben}. In
fact this linear regime is time-bound and gives way to saturation,
while the entropy in this linear regime scales as $N^2$
$(\hbar^-2)$. This is illustrated in Fig.~\ref{qent} for the system we
consider in this paper.  {\narrowtext
\begin{figure}[h]
\epsfxsize=3.5in
\epsfysize=3in
\epsfbox{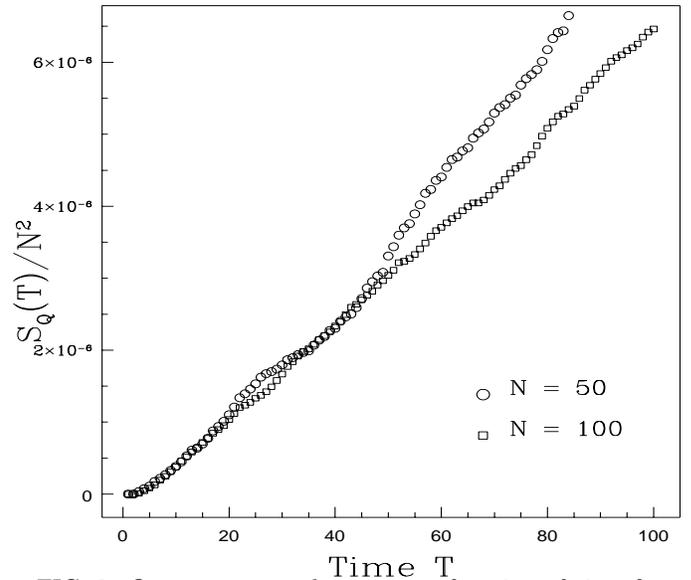}
\caption{Quantum entanglement as a function of time for
the same initial state as in Fig.~\ref{entint}. $K_1=6.0$, $K_2=5.0$,
while The interaction is $b=0.001$. The linear regime is seen to scale 
as $N^2$.}
\label{qent}
\end{figure}
}
The linear entropy increase indicates an exponentially rapid increase
in the number of single-particle basis states used. Thus we conjecture
that this regime is in fact a classical one, with this being a
reflection of the exponential increase in the number of basis states
explored in the classical Hilbert space. The rate being proportional
to the sum of positive classical Lyapunov exponents seems to suggest
that the classical entanglement production rate is proportional to the
classical K-S entropy.  The subsequent saturation is a quantum
effect, while we may expect the classical entropy to continue to
linearly increase. We further conjecture that the difference between 
chaotic and near-integrable or integrable systems will be reflected in the
classical entanglement $S_C(T)$ behaving at large times as either
$T$ or $\log(T)$, respectively. Much larger classical calculations will
be able to prove or disprove these statements.

\end{multicols}

\end{document}